\documentclass[manuscript]{acmart}
\usepackage[utf8]{inputenc}
\usepackage{bm}
\usepackage{amssymb}
\usepackage{amsmath,amsthm,amsfonts}
\usepackage{textcomp}
\usepackage{booktabs}
\usepackage{caption}
\usepackage{float}
\usepackage{multirow}

\copyrightyear{2020}
\acmYear{2020}
\setcopyright{acmcopyright}\acmConference[RecSys '20]{Fourteenth ACM Conference on Recommender Systems}{September 22--26, 2020}{Virtual Event, Brazil}
\acmBooktitle{Fourteenth ACM Conference on Recommender Systems (RecSys '20), September 22--26, 2020, Virtual Event, Brazil}
\acmPrice{15.00}
\acmDOI{10.1145/3383313.3412226}
\acmISBN{978-1-4503-7583-2/20/09}

\title{Explainable Recommendations via Attentive Multi-Persona Collaborative Filtering}
\author{Oren Barkan}
\authornote{Equal contribution.}
\email{barkanoren@gmail.com}
\affiliation{%
  \institution{Department of Computer Science, Ariel University}
  \country{Israel}
}

\author{Yonatan Fuchs}
\authornotemark[1]
\email{yonatanfuchs@mail.tau.ac.il}
\affiliation{%
  \institution{Department of Electrical Engineering, Tel Aviv University}
  \country{Israel}
}

\author{Avi Caciularu}
\email{avi.c33@gmail.com}
\affiliation{%
  \institution{Department of Computer Science, Bar-Ilan University}
  \country{Israel}
}

\author{Noam Koenigstein}
\email{noamk@tauex.tau.ac.il}
\affiliation{%
  \institution{Department of Industrial Engineering, Tel Aviv University}
  \country{Israel}
}

\begin{document}

\begin{abstract}
    Two main challenges in recommender systems are modeling users with heterogeneous taste, and providing explainable recommendations. In this paper, we propose the neural Attentive Multi-Persona Collaborative Filtering (AMP-CF) model as a unified solution for both problems. AMP-CF breaks down the user to several latent ‘personas’ (profiles) that identify and discern the different tastes and inclinations of the user. Then, the revealed personas are used to generate and explain the final recommendation list for the user. AMP-CF models users as an attentive mixture of personas, enabling a dynamic user representation that changes based on the item under consideration. We demonstrate AMP-CF on five collaborative filtering datasets from the domains of movies, music, video games and social networks. As an additional contribution, we propose a novel evaluation scheme for comparing the different items in a recommendation list based on the distance from the underlying distribution of “tastes” in the user’s historical items. Experimental results show that AMP-CF is competitive with other state-of-the-art models. Finally, we provide qualitative results to showcase the ability of AMP-CF to explain its recommendations. 
\end{abstract}



\maketitle

\section{Introduction}
Collaborative Filtering (CF) methods model users’ taste and inclinations based on user-item interactions. Often, a user is represented by a single latent vector that encodes all of the user’s different interests. For example, suppose the user likes both horror movies as well as comedies, the latent user representation should represent a positive inclination towards movies of both genres. At prediction time, both comedies and horror movies will receive high scores, but there is no easy way to distinguish between items from each genre, or explain the reason for each item recommendation. Moreover, in some cases, the more dominant genre may overtake the entire recommendation list. To mitigate this, additional diversity enhancing mechanisms are often added on top of the original CF model in order to ensure that both types of movies are properly represented within the list of recommendations \cite{1}. 

In the example above, we have used known movie genres (Comedy and Horror) to indicate two groups of user tastes or “personas”. In the general case, a user can have multiple latent personas that could not always be labeled as distinct genres. However, we assume that user interests are not monolithic and should be represented by several distinct personas, each corresponding to a different taste. This concept is somewhat related to a well-known dilemma in recommendations literature known as the accuracy-vs-diversity tradeoff \cite{29}. However, “diversity” implies that the items in the recommendation lists should be different from each other (diversified). In this work, on the other hand, we wish to identify and discern the users’ different tastes and inclinations. Our goal is to learn the correct distribution of tastes per user and the contribution or importance of each persona when different items are considered. Then, when generating a recommendation list for the user, each recommendation is associated with the user persona that best explains it.

We present Attentive Multi-Persona Collaborative Filtering (AMP-CF). AMP-CF is an explainable recommender system that models a user via an attentive mixture of personas that discerns and captures the user’s different inclinations, and explain each recommendation in the final recommendation list. Using a novel attention mechanism, the distinct user personas are dynamically weighted and combined to generate a single attentive user representation that\break depends on the candidate item (the item under consideration). When considering a specific item, some user personas may be indifferent to the item, while other personas may have more distinct positive or negative affinities. The attention mechanism in AMP-CF determines the “importance” of each persona w.r.t. the specific item and the attentive user representation dynamically updates accordingly. Thus, the AMP-CF model facilitates a type of contextual user representation: while the user personas are static at prediction time, the resulting user representation dynamically changes based on the item under consideration. Arguably, this approach better emulates the way human beings actually assess items - different characteristic components of our taste profile contribute differently in reaction to different items under consideration. Moreover, the dominant characteristics are the ones that best explain our final choices.

It is true that “traditional” single-persona user representations have the ability to encode multiple tastes through the use of orthogonal latent dimensions. However, there is no easy way to distinguish and identify the distinct tastes. In this work, we separate the traditional user representation into several latent personas, each representing a different inclination. As an additional contribution, we introduce a new evaluation procedure named Taste Distribution Distance (TDD). TDD compares the distribution of “tastes” in a recommendation list with that of the user’s interests based on her historical items.  Importantly, TDD is different than diversity. While diversity concerns with how different the recommended items are from one another, TDD measures the ability of the recommendation list to proportionally match the user’s full range of interests.

We present extensive experimental results that demonstrate the effectiveness of AMP-CF on five different CF datasets in the domains of movies, music, video games, and social networks. We show that AMP-CF outperforms state-of-the-art models both in terms of accuracy as well as in terms of TDD. Finally, we provide qualitative results that demonstrate the ability of AMP-CF to explain each recommended item based on the most relevant persona.

\section{Related Work}
Early CF models have focused on explicit ratings, where a user actively rates the items (e.g., on a 1-5 star scale) \cite{3}. Notably, learning user and item embeddings via Matrix Factorization (MF) techniques \cite{4} have gained much attention following the seminal Netflix Prize competition \cite{5}. Over the years, research have shifted towards implicit datasets, where indirect user preferences are modeled (e.g., clicks or a purchase) \cite{6,7}.

Neural embedding research \cite{23,34,35,36,37,38,39,40,41} has been a catalyst for new approaches in recommender systems. Salakhutdinov et al. \cite{8} have pioneered Restricted Boltzman Machines (RBMs) for CF. In More recently, autoencoders have been popularized for recommender systems \cite{9,10,11}. He et al. \cite{2} proposed the popular Neural Collaborative Filtering model (NCF, also termed NeuMF). NCF employs a fusion between Generalized Matrix Factorization (GMF) and Multilayer Perceptron (MLP) for predicting the user-item relations. This model was also shown to outperform many strong CF baselines such as eALS \cite{12}, BPR \cite{13}, ItemKNN \cite{3}, and recently even memory-network models such as CMN \cite{15}. 

Attentive recommendation models have been described in several earlier works such as Attentive CF \cite{14}, Neural Attentive Multiview Machines \cite{32}, DeepICF \cite{33}, and Attentive Item2vec \cite{23}. These works model users as a set of item vectors. In contrast, a unique property of AMP-CF is its ability to represent a user by a unique set of personas.  The user personas are weighted w.r.t to the specific target item under consideration in order to dynamically produce the final user representation. 

Additional line of research is concerned with learning a sequence of user activities. For example, Convolutional Sequence Embedding Recommendation (Caser) \cite{20} is an algorithm for modelling user sessions. In other cases, temporal data is utilized to predict the next item based on recent activity. For example, Distance Based Memory Recommender (SDMR) \cite{30} is not a sequential model, but it does utilize recent activities as contextual side-information. The contribution and the focus of this work is different from these models, as AMP-CF does not utilize any temporal information. Hence, we do not evaluate against these models and leave the entire question of temporal data for future work. 

\section{Attentive Multi-persona Collaborative Filtering}
\label{sec:AMP-CF}
In this section, we describe the AMP-CF model in detail. Let $\mathcal{I}=\{i\}_{i=1}^{N_u}$ and $\mathcal{J}=\{j\}_{j=1}^{N_v}$ be the sets of identifiers corresponding to the users and items, respectively. An item $j$ is represented by a latent vector $v^j\in \mathbb{R}^d$. A user $i$ is represented by a multi-persona matrix $U^i\in \mathbb{R}^{r\times d}$. The $k$-th row in $U^i$, denoted by $u_k^i$, represents the $k$-th persona of user $i$.

\begin{figure*}[tb]
\centering
\includegraphics[width=1\linewidth]{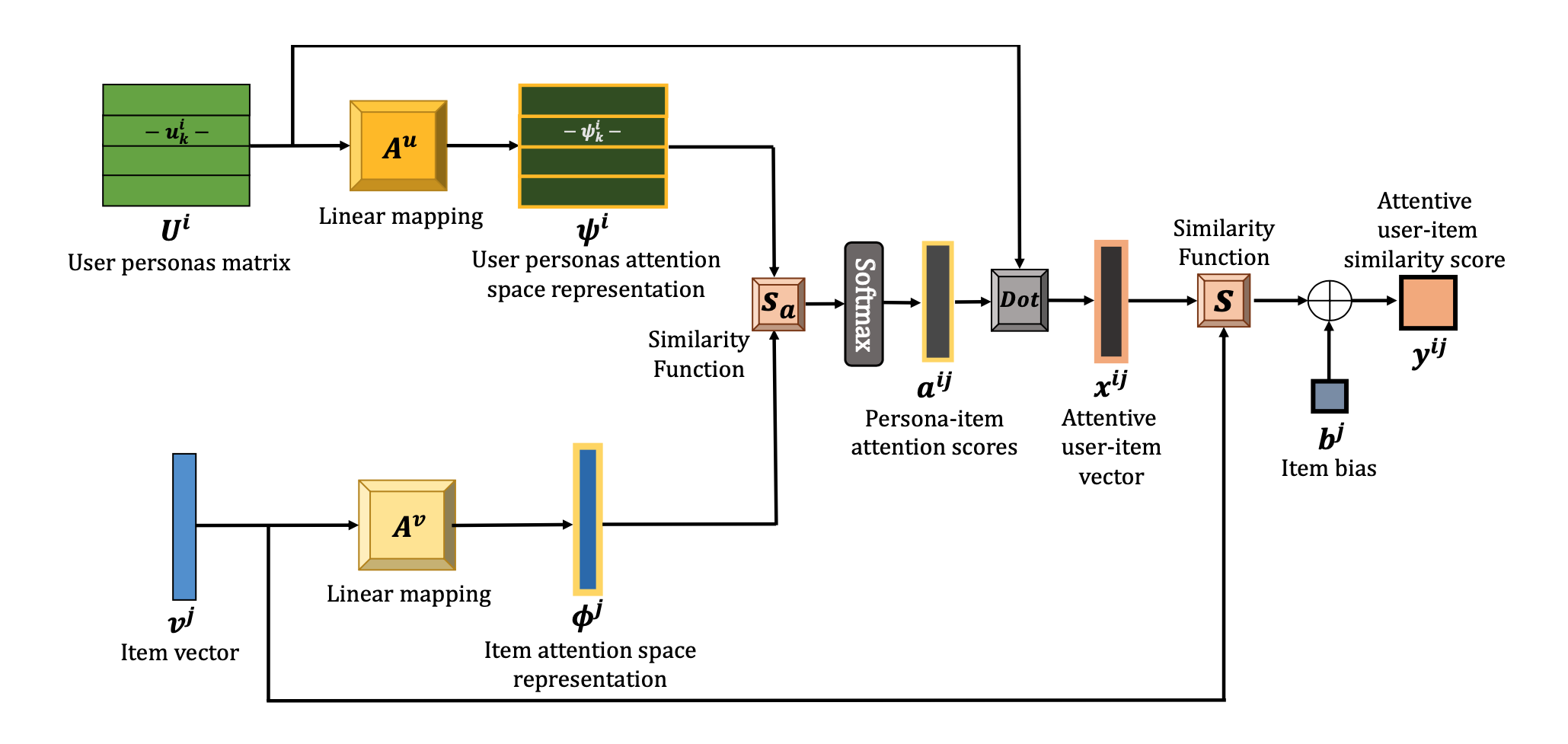}
\caption{The AMP-CF model architecture. See Sec.~\ref{sec:AMP-CF} for details.}
\label{fig:model_1}
\end{figure*}

Let $A^u\in \mathbb{R}^{d\times d_a}$ and $A^v\in\mathbb{R}^{d_a\times d}$ be linear mappings from the user and item original feature space to the user-item attention space, and denote $\psi_k^i\triangleq u_k^i A^u$ and $\phi^j\triangleq A^v$, and $v^j$ as the user persona and item representations in the attention space, respectively. In this space, the affinities between each of the user personas and an item are computed via a general similarity function $s_a: \mathbb{R}^{d_a} \times\mathbb{R}^{d_a }\rightarrow\mathbb{R}$. We denote the affinity between the $k$-th persona of user $i$ and the item $j$ by $s_a^{ikj}\triangleq s_a (\psi_k^i,\phi^j)$. These affinities are used to form an attentive user-item vector $x^{ij}$, which is a convex combination of the user’s personas as follows: $x^{ij}=U^ia^{ij}=\sum_{k=1}^r a^{ij}_k u^{i}_k$, with $a_{k}^{ij}=\frac{\exp(s_a^{ikj})}{\sum_{m=1}^r \exp(s_a^{imj})}$. Finally, the attentive user-item similarity score $y^{ij}$ is computed as follows:
\begin{equation}
y^{ij}=s^{ij}+b^{j},
    \label{eq:similarity}
\end{equation}
where $s: \mathbb{R}^d\times \mathbb{R}^d\rightarrow\mathbb{R}$ is a general similarity function, $s^{ij}\triangleq s(x^{ij},v^j)$, and $b^j$ is the bias for the item $j$. The AMP-CF model is depicted in Fig.~\ref{fig:model_1}. For simplicity, in our implementation, we set both $s$ and $s_a$ to the dot-product. Alternatively, these similarity functions can be set to small neural networks with non-linear activations \cite{16,17}. However in our experimentation, this failed to yield any additional improvement, while increasing training times.

\subsection{AMP-CF Training and Inference}
AMP-CF is trained using a dataset of implicit user-item relations. We assume that for each user we have a set of items that she consumed. We treat these user-item pairs as positive relations and they define the training set $\mathcal{D}=\{(i,j) | \text{ user } i \text{ consumed item } j\}$. We also denote $\mathcal{N}\subset\mathcal{J}$ - a subset of items that are sampled (stochastically, during the training phase, for each positive pair $(i,j)\in\mathcal{D}$) from $\mathcal{J}$ according to the unigram item distribution raised to the power of $0.5$ \cite{18}. The items in $\mathcal{N}$ are treated as negative target items w.r.t. the user $i$. In order to encourage each persona to acquire a different taste, we calculate the entropy of the attentive scores on both the positive and negative samples and incorporate them into the objective function. That is, for each user-item interaction $(i,j)$ our loss $\mathcal{L}$ consists of the weighted sum of the negative log-likelihood loss $\mathcal{L}_\mathcal{D}^{ij}$, and the entropy based loss $\mathcal{L}_\mathcal{H}^{ij}$ as follows:


\begin{align}
\label{eq:loss}
\quad\quad\mathcal{L}=\sum_{(i,j)\in \mathcal{D}} \alpha \mathcal{L}_\mathcal{D}^{ij}+(1-\alpha)\mathcal{L}_\mathcal{H}^{ij},
\end{align}
where
\begin{align*}
\qquad\qquad\qquad\quad\mathcal{L}_\mathcal{D}^{ij}=&-\log p(j|i)=-y^{ij}+\log [ \sum_{n\in\mathcal{N}\cup\{j\}} \exp(y^{in})],\\
\mathcal{L}_\mathcal{H}^{ij}=&\lambda_p\mathcal{H}_p^{ij}-\lambda_n\mathcal{H}_n^{ij},
\end{align*}
and
\begin{align*}
\mathcal{H}_p^{ij}=&-\sum_{k=1}^r a_k^{ij} \log a_k^{ij},\\
\mathcal{H}_n^{ij}=&-\sum_{n\in \mathcal{N}} \sum_{k=1}^r a_k^{in} \log a_k^{in}.
\end{align*}


The objective in Eq. \ref{eq:loss} can be optimized via any stochastic gradient descent method. The entropy loss $\mathcal{L}_\mathcal{H}^{ij}$ ensures that the resulting user personas are different from each other and efficiently  utilized to capture the users’ distinct tastes and inclinations. The minimization of the positive entropy $\mathcal{H}_p^{ij}$ causes the model to shift the full attention to only one persona for each positive example, thus helping it to acquire a specific taste. On the other hand, the the negative examples are randomly sampled and should not be attributed to any specific persona in particular.  Hence, the
minimization of negative entropy $-\mathcal{H}_n^{ij}$ is required in order to ensure that all personas take an equal part in scoring the negative examples. $\alpha$ is a hyperparameter that controls the tradeoff between $\mathcal{L}_\mathcal{H}^{ij}$ and $\mathcal{L}_\mathcal{D}^{ij}$. Finally, at the inference phase, AMP-CF uses $y^{ij}$ (Eq. \ref{eq:similarity}) to score the similarity between user $i$ and item $j$.

\begin{table*}[]
\caption{Dataset Statistics}
\label{tab:tab1}
\begin{tabular}{@{}lcccccccccc@{}}
\toprule
\multicolumn{1}{c}{\multirow{2}{*}{Dataset}} & \multirow{2}{*}{$\#$ Users} & \multirow{2}{*}{$\#$ Items} & \multirow{2}{*}{$\#$ Training Examples} & \multirow{2}{*}{Sparsity} & \multicolumn{3}{c}{$\#$ Ratings Per User} & \multicolumn{3}{c}{$\#$ Ratings Per Item} \\ \cmidrule(l){6-11} 
\multicolumn{1}{c}{}                         &                          &                          &                                      &                           & Mean         & Std         & Med       & Mean         & Std         & Med       \\ \midrule
ML-100k                                      & 610                      & 6278                     & 96683                                & 0.975                     & 158.6        & 250.9       & 69        & 15.4         & 26.2        & 6         \\ \midrule
ML-1M                                        & 6040                     & 3592                     & 993060                               & 0.954                     & 164.5        & 192.7       & 95        & 276.7        & 384.1       & 130       \\ \midrule
Yahoo!                                       & 14990                    & 5394                     & 397733                               & 0.995                     & 26.5         & 30.5        & 17        & 73.8         & 259.5       & 6         \\ \midrule
Amazon                                       & 7868                     & 12802                    & 98421                                & 0.999                     & 12.5         & 11.8        & 9         & 7.6          & 12.8        & 4         \\ \midrule
Pinterest                                    & 15001                    & 9175                     & 374017                               & 0.997                     & 24.9         & 7.3         & 23        & 40.8         & 40.6        & 29        \\ \bottomrule
\end{tabular}
\end{table*}

\section{Experiment Setup and Results}
In this section, we describe the evaluation process, the datasets, data preparation, hyperparameter configuration, and present the results.
\subsection{Datasets and Data Preparation}
Five public datasets are considered. Dataset statistics are presented in Tab. 1. We filtered and processed these datasets as follows:

{\bf MovieLens 100K (ML100K)}: This is a small version of the MovieLens 20M dataset \cite{25} containing 100K user-movie ratings using a 5-star scale (0.5 - 5.0). We considered all items as positive examples and filtered users with less than two items. 

{\bf MovieLens 1M (ML1M)}: This is another version of the MovieLens 20M dataset \cite{25} containing 1M user-item ratings. The preprocessing of this dataset was done in the same manner as the ML100K dataset.

{\bf Pinterest:} This implicit feedback dataset \cite{26} was constructed by \cite{2} for evaluating content-based image recommendation. We followed \cite{2} and constructed the training set such that each user has at least 20 positive interactions.

{\bf Amazon Video Games:} This dataset \cite{27} contains Amazon video games ratings on a 5-star scale ($1.0$ – $5.0$). The preparation of the data was done in the same manner as the ML100K dataset.

{\bf Yahoo! Music:} This dataset \cite{28} contains user-artist ratings on a 100-score integer scale ($0$ – $100$), with a special score of 255 given to songs that were extremely disliked. The preparation of the data was done in the same manner as the ML100K dataset.

\subsection{Evaluation Measures and Protocols}
\label{subsec:eval_meas}
The first part of our evaluations share a similar experimental setup as in previous works \cite{2,15}: We reserve the last item of each user for the test set, and the rest of the items are used for training the user personas and item representations. We compute and report the Hit Ratio at 10 (\textbf{HR@10}) and Normalized Discounted Cumulative Gain at 10 (\textbf{NDCG@10}) measures \cite{29} averaged across users. 

In the second part of our evaluation, we care about identifying users’ distinct tastes and inclinations. To this end, we introduce a new evaluation procedure named Taste Distribution Distance (TDD). The idea behind TDD is to generate a “neutral” similarity space, independent of any of the evaluated models. Based on this similarity space, we wish to encode each user’s different tastes as a distribution induced by the user’s historical items and compare it to a distribution induced by the recommendation lists generated by each model under evaluation. Ideally, the distribution of tastes in the recommendation list will match the one induced by the user’s historical items.

Let $n$ and $m$ be the number of users and items, respectively. In order to compute TDD, we first form a binary interaction matrix $R\in \mathbb{R}^{m\times n}$, where each entry $R_{ij}$ is equal to $1$ if the user $i$ consumed the item $j$ otherwise $0$. Hence, the columns of $R$ represent items based on the users that consumed them. Next, we apply Principal Component Analysis (PCA) to reduce the dimension of the item vectors to 100, s.t. each item $j\in J$ is represented by a PCA vector $r_j\in\mathbb{R}^{100}$. Next, we applied K-Means clustering to group the items into $K = 50$ different clusters representing 50 distinct latent tastes.  The TDD metric is calculated as follows: First, for each user $i$ we compute a list of the top 30 item recommendations produced by the model. Then, for each item $j$ from $i$’s recommendation list, a vector $d_{ij}\in\mathbb{R}^{50}$ is calculated by taking the cosine distance between the item and each of the 50 clusters’ means. We average the $d_{ij}$ vectors across user $i$’s recommendation list and apply a softmax function to produce a distribution vector ${\widetilde{d}}_i$. Hence, ${\widetilde{d}}_i$ represents the distribution of tastes induced by $i$’s recommendation list over the 50-dimensional taste space. Next, the same process is applied to user $i$’s historical items to produce the distribution vector ${\widetilde{t}}_i$. Hence, ${\widetilde{t}}_i$ represents a distribution over $i$’s different tastes induced by her historical items. Finally, the difference between ${\widetilde{d}}_i$ and ${\widetilde{t}}_i$ is measured based on the Jensen–Shannon divergence and the Hellinger distance, defined respectively as follows: $JS\left({\widetilde{d}}_i,{\widetilde{t}}_i\right)=\sqrt{\frac{D_{KL}({\widetilde{d}}_i\parallel m_i)+D_{KL}({\widetilde{t}}_i\parallel m_i)}{2}}$, and $Hel\left({\widetilde{d}}_i,{\widetilde{t}}_i\right)=\frac{1}{\sqrt2}\bigg\|{\sqrt{\widetilde{d}}_i}-{\sqrt{{\widetilde{t}}_i}}\bigg\|_2$, where $m_i$ is the pointwise mean of $\widetilde{d_i}$ and $\widetilde{t_i}$, and $D_{KL}$ is the Kullback-Leibler divergence.

Importantly, the TDD procedure is different than simply measuring the diversity in the recommendation list. While simple diversity measures the extent in which the different items in the recommendation list are different from each other, the goal of TDD is to measure the extent in which the recommendation list corresponds with the different user tastes as reflected by the user’s historical items. For a user with multiple distinct tastes, TDD requires that the user’s recommendation list will cover all her tastes at the same proportions as in her historical items. On the other hand, for a user with a single taste, TDD prohibits diversification and requires that all recommendations will follow her one single interest. We consider TDD as an additional contribution of this work.

\subsection{Evaluated Models} 
The following models participate in the evaluation:

{\bf AMP}: The proposed model from Sec.~\ref{sec:AMP-CF}. We consider several versions of AMP-CF varying by the number of personas$ r\in{2,\ 4,\ 8}$, e.g., for $r=2$, we denote AMP-2. We set $d=d_a=64$ for all AMP-CF variants.

{\bf DeepICF+a}: Deep Item Collaborative Filtering \cite{33}. This state-of-the-art model uses a deep neural network to learn higher-order interactions and assign attention weights to the target items according to every item in a user’s training history. Following \cite{33}, we used the pretraining regime using FISM, and a latent dimension of $d=64$.NCF: Neural Collaborative Filtering (NCF) \cite{2} is a well-known model that was shown to significantly outperform strong baselines such as BPR \cite{13}, eALS \cite{12} and ItemKNN \cite{3}, and recently even memory based networks such as CMN \cite{15}. Therefore, and due to scope limitations, we do not further report results for these models and compare our model to NCF. Following \cite{2}, we used a latent feature dimension of d=64 that was shown to produce the best results (we further tried 8, 16, 32, 64, and 128).

{\bf ConvNCF}: Convolutional Neural Collaborative Filtering (ConvNCF) \cite{19} is a state-of-the-art multi-layer convolutional neural network model which uses an outer product to explicitly model the pairwise correlations between the dimensions of the embedding space. The resulting interaction map is passed through a convolutional neural network and trained with BPR loss. In accordance with \cite{19}, we trained the network with $d=64$ for the users and items embeddings and six layers in the neural network.

{\bf DiRec}: DiRec \cite{22,24} is a novel technique for diversifying recommendations. Items are clustered together in priority-medoids and cover-trees are used to declaratively capture a desired balance between predicted rating and diversity. In \cite{23}, DiRec was compared to other algorithms such as Greedy and Swap \cite{31} and showed to significantly outperformed them on several evaluation metrics. 

{\bf AISP}: Considering the TDD evaluation defined above, one may wonder whether it will suffice to cluster the PCA vectors from the TDD space directly in order to maximize the TDD metrics. In order to disproof this proposition and to further showcase the contribution of our model, we propose the Attentive Interaction-Space Personas (AISP) baseline. In AISP, PCA item vectors are extracted according to the description in Sec.~\ref{subsec:eval_meas}. Then, for each user, we compute K-Means over her training items to produce K=p means that serve as an alternative to the user personas. At inference, we use the AMP-CF pipeline (Sec.~\ref{sec:AMP-CF}) to produce attentive user-item scores. However, in AISP, both $A^u$ and $A^v$, and $s$ and $s_a$, are not learned. Instead, we set $s$ and $s_a$ to be simple dot-product and set $A^u=A^v=I$.

For AMP-CF, NCF and ConvNCF we used negative sampling parameter $\left|\mathcal{N}\right|=4$ (when applied) - other values (10, 20, 100) were found to perform worse. AMP-CF and NCF Models were optimized using Adam \cite{21} with learning rate 0.001 and minibatch size of 256, which is the best configuration from \cite{2}. ConvNCF was optimized according to the process described in \cite{19}. Optimization proceeds until a saturation in HR@10 and NDCG@10 is reached over the validation set (early stopping procedure).

\begin{table*}[t]
\caption{HR@10 and NDCG@10 for all models and datasets.}\vspace*{-8pt}
\label{tab:tab2}
\begin{tabular}{@{}lcccccccccc@{}}
\toprule
          & \multicolumn{5}{c}{HR@10}                                                               & \multicolumn{5}{c}{NDCG@10}                                                             \\ \midrule
          & ML100K          & ML1M            & Pinterest       & Yahoo!          & Amazon          & ML100K          & ML1M            & Pinterest       & Yahoo!          & Amazon          \\ \midrule
DeepICF+a & 0.7314          & 0.7342          & 0.7952          & 0.9218          & 0.5608          & 0.4801          & 0.4690          & 0.5011          & 0.6985          & 0.3748          \\ \midrule
NCF       & 0.7317          & 0.7313          & 0.7948          & 0.9174          & 0.5657          & 0.4786          & 0.4694          & 0.5026          & 0.6984          & 0.3750          \\ \midrule
ConvNCF   & 0.7138          & 0.7129          & 0.6487          & 0.8917          & 0.3456          & 0.4532          & 0.4470          & 0.3911          & 0.6312          & 0.1946          \\ \midrule
DiRec     & 0.6132          & 0.5516          & 0.7767          & 0.9091          & 0.5512          & 0.4122          & 0.3434          & 0.4891          & 0.6754          & 0.3694          \\ \midrule
AISP-2    & 0.6895          & 0.6189          & 0.7937          & 0.9211          & 0.5752          & 0.4620          & 0.3742          & 0.5030          & 0.7001          & 0.3824          \\ \midrule
AISP-3    & 0.6896          & 0.6207          & 0.7814          & 0.9111          & 0.5713          & 0.4588          & 0.3750          & 0.4996          & 0.6974          & 0.3829          \\ \midrule
AISP-4    & 0.6908          & 0.6219          & 0.777           & 0.9176          & 0.5708          & 0.4598          & 0.3754          & 0.4928          & 0.6844          & 0.4058          \\ \midrule
AMP-2     & \textbf{0.7376} & \textbf{0.7358} & \textbf{0.7966} & \textbf{0.9236} & 0.5751          & \textbf{0.4935} & \textbf{0.4706} & \textbf{0.5031} & 0.6988          & \textbf{0.3888} \\ \midrule
AMP-3     & 0.7244          & 0.7316          & 0.7893          & 0.9229          & \textbf{0.5788} & 0.4813          & 0.4653          & 0.4972          & \textbf{0.7005} & 0.3858          \\ \midrule
AMP-4     & 0.7302          & 0.7280          & 0.7963          & 0.9235          & 0.5750          & 0.4855          & 0.4601          & 0.5017          & 0.6997          & 0.3855          \\ \bottomrule
\end{tabular}
 \end{table*}

 \begin{table*}[t]
\caption{JS Divergence and Hellinger Distance for all models and datasets.}\vspace*{-8pt}
\label{tab:tab3}
\begin{tabular}{@{}lllllllllll@{}}
\toprule
          & \multicolumn{5}{c}{JS Divergence}                                                                                                               & \multicolumn{5}{c}{Hellinger Distance}                                                                                                          \\ \midrule
          & \multicolumn{1}{c}{ML100K} & \multicolumn{1}{c}{ML1M} & \multicolumn{1}{c}{Pinterest} & \multicolumn{1}{c}{Yahoo!} & \multicolumn{1}{c}{Amazon} & \multicolumn{1}{c}{ML100K} & \multicolumn{1}{c}{ML1M} & \multicolumn{1}{c}{Pinterest} & \multicolumn{1}{c}{Yahoo!} & \multicolumn{1}{c}{Amazon} \\ \midrule
DeepICF+a & 1.2641                     & 1.1089                   & 1.4526                        & 1.5001                     & 1.7892                     & 0.0243                     & 0.0203                   & 0.0551                        & 0.0416                     & 0.0686                     \\ \midrule
NCF       & 1.3939                     & 1.1568                   & 1.5749                        & 1.5728                     & 1.9596                     & 0.0332                     & 0.0206                   & 0.0531                        & 0.0447                     & 0.0711                     \\ \midrule
ConvNCF   & 2.2429                     & 1.2977                   & 1.8289                        & 2.0377                     & 2.6290                     & 0.0818                     & 0.0266                   & 0.0806                        & 0.0773                     & 0.1577                     \\ \midrule
DiRec     & 1.7340                     & 1.2687                   & 2.0426                        & 1.8531                     & 2.1731                     & 0.0493                     & 0.0212                   & 0.1050                        & 0.0519                     & 0.0895                     \\ \midrule
AISP-2    & 2.2308                     & 1.5839                   & 1.5458                        & 1.5755                     & 1.6305                     & 0.0776                     & 0.0389                   & 0.0496                        & 0.0414                     & 0.0496                     \\ \midrule
AISP-3    & 2.1929                     & 1.5645                   & 1.6825                        & 1.5611                     & 1.6401                     & 0.0747                     & 0.0380                   & 0.0605                        & 0.0454                     & 0.0487                     \\ \midrule
AISP-4    & 2.1834                     & 1.5558                   & 1.8033                        & 1.5609                     & 1.6487                     & 0.0740                     & 0.0375                   & 0.0716                        & 0.0413                     & 0.0493                     \\ \midrule
AMP-2     & 1.2390                     & 1.1085                   & 1.3800                        & 1.4520                     & \textbf{1.5825}            & 0.0265                     & 0.0191                   & 0.0399                        & 0.0407                     & 0.0452                     \\ \midrule
AMP-3     & 1.1611                     & \textbf{1.0930}          & 1.4114                        & \textbf{1.4279}            & 1.6450                     & 0.0231                     & \textbf{0.0186}          & 0.0403                        & \textbf{0.0388}            & 0.0480                     \\ \midrule
AMP-4     & \textbf{1.1062}            & 1.1169                   & \textbf{1.3305}               & 1.4446                     & 1.6007                     & \textbf{0.0213}            & 0.0194                   & \textbf{0.0365}               & 0.0400                     & \textbf{0.0437}            \\ \bottomrule
\end{tabular}
\end{table*}

\subsection{Results}

Table ~\ref{tab:tab2} presents the HR@10 and NDCG@10 for all combinations of models and datasets. Similarly, Tab. ~\ref{tab:tab3} presents the Jensen-Shannon Divergence and Hellinger Distance TDD metrics respectively. We see that the attribute this to the fact that in more complex domains such as Music or Pinterest images, people are more likely to have multiple distinct interests. In these cases, AMP-CF better accommodates the distance between item-vectors, as the entropy-loss pushes personas at different directions during training. On the other hand, the Amazon dataset demonstrates relatively low TDD gains. With very fewer items per user (see the statistics in Tab. 1), this dataset is less likely to have users with diversified taste. 

Next, we turn to the AISP baseline. Recall, this baseline is based on the same PCA vectors that were used for TDD. Nevertheless, the AMP-CF model outperforms the AISP baseline in both TDD metrics. This serves to disproof the notion that clustering the raw PCA vectors will suffice to accurately discern the users interests and achieve good TDD measures. Moreover, it showcases the ability of AMP-CF to emulate the correct taste distribution at the recommendation phase even when compared to a competitive baseline such as AISP. 

\begin{figure*}[h] 
\centering
\includegraphics[width=1\linewidth]{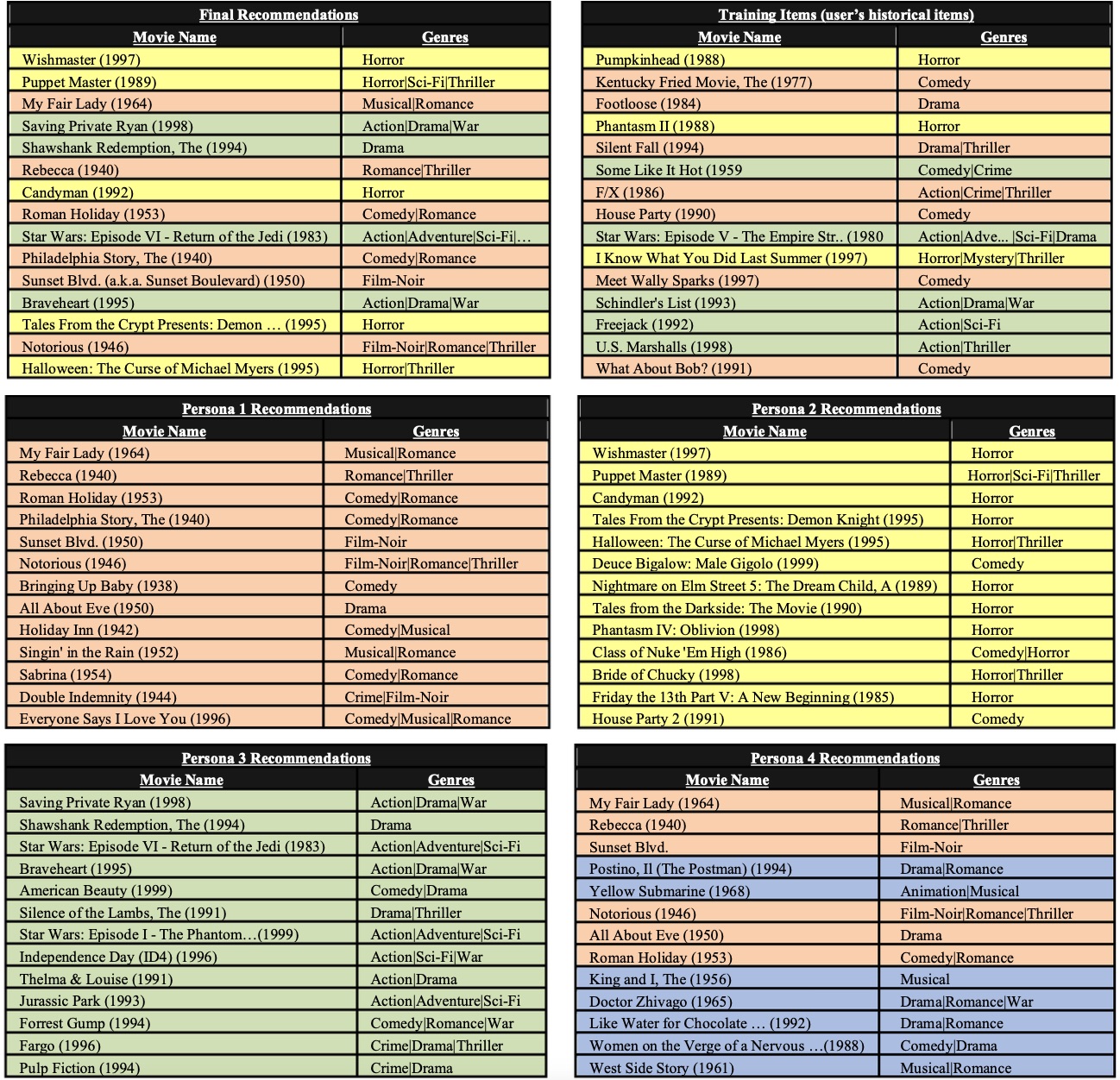}
\caption{Explainable recommendations using AMP-CF. The figure presents the historical (training) items of a randomly selected user, and the recommendations list together with a breakdown into four personas, produced by AMP-CF (AMP-4). Personas 1, 2, and 3 are marked in red, yellow and green, respectively. Persona 4, which converged quite similar to Persona 1, is marked in both red (overlapping items with persona 1) and blue. The final recommendation list is a mixture of the internal recommendations of the different personas, where each item recommendation is explained by a specific persona. The items in the training items list are color-coded according to the persona which had the highest attentive score for the item at the end of the training phase.}
\label{fig:fig2}
\end{figure*}

Finally, Fig. ~\ref{fig:fig2} showcases the ability of AMP-CF to learn distinctive personas in terms of the user’s tastes. To this end, we chose a random user and inspected her training items, final recommendations list, and per-persona recommendations produced using a four-persona model (AMP-4). The recommendations of each individual persona were produced by choosing the top items w.r.t. to each persona vector in separate. As can be seen, three of the personas acquired three distinct tastes (marked by three different colors) whilst the fourth converged somewhere in the middle between the others, hence, it is marked by two colors: red for
recommendations that overlaps with persona 1, and blue for the rest. Each training item was colored according to the persona which produced the highest attentive score for that item. The final recommendation list is a mixture based on each persona’s list, sorted by overall score . This qualitative example demonstrates the ability of AMP-CF to learn latent personas that reveal the different user’s tastes and explain the recommendations in the final recommendation list.

\section{Conclusion}

This paper presents AMP-CF – a neural attentive model that represents each user as an attentive mixture of personas. AMP-CF discerns user interests into separate personas and facilitates a novel attention mechanism to determine the importance of each persona w.r.t. the item under consideration. Then, AMP-CF dynamically produces an attentive user-item representation that represents the user inclinations when considering the specific item. The learned personas serve as a source of explanation for the items that appear in the final recommendation list for the user. In addition, we proposed the taste distribution distance (TDD) analysis that provides a sensitive framework for assessing the ability of recommendations algorithms to emulate users’ taste distribution. Extensive experimental results on five different datasets demonstrate the effectiveness of AMP-CF when compared to different state-of-the-art baselines. Last but not least, our qualitative analysis showcases the ability of AMP-CF to reveal the user’s latent personas and facilitate an explainable recommender system.

\begin{acks}
This research was supported by the ISRAEL SCIENCE FOUNDATION (grant No. 2243/20)
\end{acks}

\bibliography{references}
\bibliographystyle{ACM-Reference-Format}

\end{document}